\documentclass[a4paper]{emulateapj}
  
\usepackage{amsmath,amssymb,amsfonts}
\usepackage{color}

\usepackage[T1]{fontenc}
\usepackage{txfonts}
\usepackage[]{graphicx}

\def\na{New Astronomy}
\def\pr{Physical Review}

\shorttitle{A Shining Death of Unequal SMBHBs}
\shortauthors{Xian Chen, D. N. C. Lin, \& Xiaojia Zhang, 2019}

\begin{document}

\title{A Shining Death of Unequal Supermassive Black Hole Binaries}

\author{Xian Chen}
\email{xian.chen@pku.edu.cn}
\affiliation{Astronomy Department, School of Physics, Peking University, 100871 Beijing, China}
\affiliation{Kavli Institute for Astronomy and Astrophysics at Peking University, 100871 Beijing, China}

\author{D. N. C. Lin}
\affiliation{Department of Astronomy and Astrophysics, University of California, Santa Cruz, CA 95064, USA}
\affiliation{Institute for Advanced Studies, Tsinghua University, Beijing, China}

\author{Xiaojia Zhang}
\affiliation{Department of Earth Sciences, University of Hong Kong, Hong Kong, China}
\affiliation{Planetary Environment and Asteroid Resource Laboratory, Origin Space Technology Co. Ltd.}

\begin{abstract}
In the $\Lambda$CDM scenario, small galaxies merge to produce larger entities.
	Since supermassive black holes (SMBHs) are found in galaxies of all
	sizes, SMBH binaries (SMBHBs) are generally expected to form during the
	amalgamation of galaxies. It is unclear what fraction of these binaries
	could eventually merge, but a general consensus is that initially the
	orbital decay is mediated by the surrounding gas and stars.  In this
	Letter, we show that in active galactic nulcei (AGNs) the radiation
	field also causes the orbits of the accreting SMBHs to shrink.  The
	corresponding mechanism, known as the ``Poynting-Robertson drag'' (PR
	drag), takes effect on a well-defined timescale $CT_{\rm Sal}$, where
	$T_{\rm Sal}$ is the Salpeter timescale of the AGN, presumably coincide
	with the primary SMBH, and
	$C=4\xi^{-1}\epsilon^{-1}q^{1/3}(1+q)^{2/3}(1-\epsilon)$ is a constant
	determined by the radiative efficiency $\epsilon$, the mass ratio $q$
	of the two black holes, and a parameter $\xi$ characterizing the size
	of the circum-secondary accretion disk. We find that when $q\la$a
	few$\times10^{-5}$, the PR drag is more efficient in shrinking the
	binary than many other mechanisms, such as dynamical friction and
	type-I migration. Our finding points to a possible new channel for the
	coalescence of unequal SMBHBs and the clearing of intermediate-massive
black holes in AGNs.  \end{abstract}

\keywords{Astrodynamics --- Quasars --- Gravitational waves--- Active galactic nuclei --- Accretion}

\section{Introduction}

Almost all massive galaxies contain supermassive black holes (SMBHs) in their
centers \citep{kormendy13}. The current consensus is that such black holes (BHs)
form in small galaxies in the early universe and grow to
$10^6-10^{10}\,M_\odot$ by episodic accretion of gas \citep{soltan82,yu02}.
These accretion phase, as is understood today, can be triggered by galaxy
mergers \citep{kauffmann00}.  During such a phases, a fraction of the
gravitational energy of the gas is released in the form of radiation and the
galaxy center becomes an active galactic nucleus (AGN). The luminosity could
exceed the Eddington limit in the most extreme case
\citep[e.g.][]{wu15}.
 
Such a close relationship between galaxy merger and SMBH growth results in an
inevitable consequence that pairs of SMBHs form in the nuclei of merging
galaxies \citep{begelman80}. Such SMBH binaries (SMBHBs) are important
astrophysical objects in the era of gravitational-wave astronomy. Merging
SMBHBs are the major targets of the ongoing Pulsar Timing Arrays
\citep{hobbs10} and the planned Laser Interferometer Space Antenna
\citep{amaro-seoane17} 

However, there is a long-standing debate regarding the coalescence of SMBHBs.
Earlier analysis of the dynamical evolution of the binaries revealed a
bottleneck when the binaries shrink to a size of about one parsec
\citep{begelman80}.  At this stage, the binaries become ``hard'' and start to
slingshot the surrounding stars. In the simplest galaxy model, i.e., the
stellar distribution is spherically symmetric and there is no gas, the
replenishing of stars to the vicinity of the SMBHBs is inefficient
\citep{makino97,quinlan97,milosavljevic01}. Consequently, the evolution of the
binaries may stall. This theoretical prediction is inconsistent with the
apparent scarcity of SMBHBs in galactic nuclei \citep{komossa06} and the
conundrum is called ``the final parsec problem''.

Real galaxies are more complicated than the idealized stellar systems based on
spherical models.  In general, 
merging galaxies are asymmetric. As a result, stars could be fed to
the galaxy centers more efficiently so that the final parsec barrier may be
avoided \citep{zhao02,yu02,merritt04,berczik06}.  Moreover, since galaxy mergers
often trigger gas inflow, it was realized early on that many SMBHBs
may reside in gaseous environments and the interaction with gas may offer a
potential solution to the above problem \citep{begelman80,ivanov99,gould00}.  

Later numerical simulations of a smaller (secondary) BH embedded in the
accretion disk of a bigger (primary) SMBH generally confirm the above picture
and, furthermore, showed that the evolution is similar to the migration of a
planet in a protoplanetary disk: The secondary opens a gap in the disk and
migrates towards the primary on the viscous timescale
\citep{armitage02,cuadra09}. Equal-mass binaries could even clear out an cavity
in the accretion disk and in this case the merger is driven mainly by the
spiral arms in the circum-binary disk \citep{macfadyen08}. In hotter environments where
the gas distribution is more or less isotropic, the SMBHBs could excite
elongated structures \citep{escala04,dotti05} or density wakes \citep{kim08}
which lag behind the major axes of the binaries. These structures impose a
negative torque on the binaries, which could also accelerate the shrinking of
the binary orbit.

However, recent hydrodynamical simulations revealed a more controversial
picture.  They show that the aforementioned gap or cavity are not empty but
filled with gas streams, which originate from the inner edge of the
circum-binary disk and end up on both BHs \citep{hayasaki07,farris14}. 
These gas streams could exert a
positive torque \citep{roedig12} as well as directly deposit angular momentum
onto the BHs \citep{hayasaki09,shi12}.  As a result, the binary orbit may even
expand so that the final parsec problem remains
\citep{miranda17,moody19,munoz19}. It is worth noting that the generality of
the expansion of the binary orbit deserves further investigation because the
evolution is sensitive to the viscosity and the thermodynamical properties of
the gas inside the gap and cavity \citep{tang17}.

So far the models have ignored the impact of the radiation of the accretion
disks on the evolution of the binaries. It is known that our Sun could induce a
drag force on the dust particles in the solar system. The drag effect is caused
by the asymmetry between the absorption and re-emission of the solar
irradiation \citep{poynting1903}. A fully relativistic treatment of the
phenomenon further clarified that the drag force can be attributed to the light
beaming effect: More light is re-emitted in the direction of motion of the dust
particles \citep{robertson1937}. Such an effect, also known as the
``Poynting-Robertson effect'' (PR effect, hereafter), in principle also applies
to a SMBHB system because (i) one of the SMBHs, by accreting gas, could radiate
and (ii) the other BH surrounded by its own accretion disk could absorb and
re-emit this radiation. Here we study this effect and show that the drag force
is indeed important for the orbital decay of SMBHBs.

\section{The Poynting-Robertson Drag}

\begin{figure}
\begin{center}
\includegraphics[clip,width=0.45\textwidth]{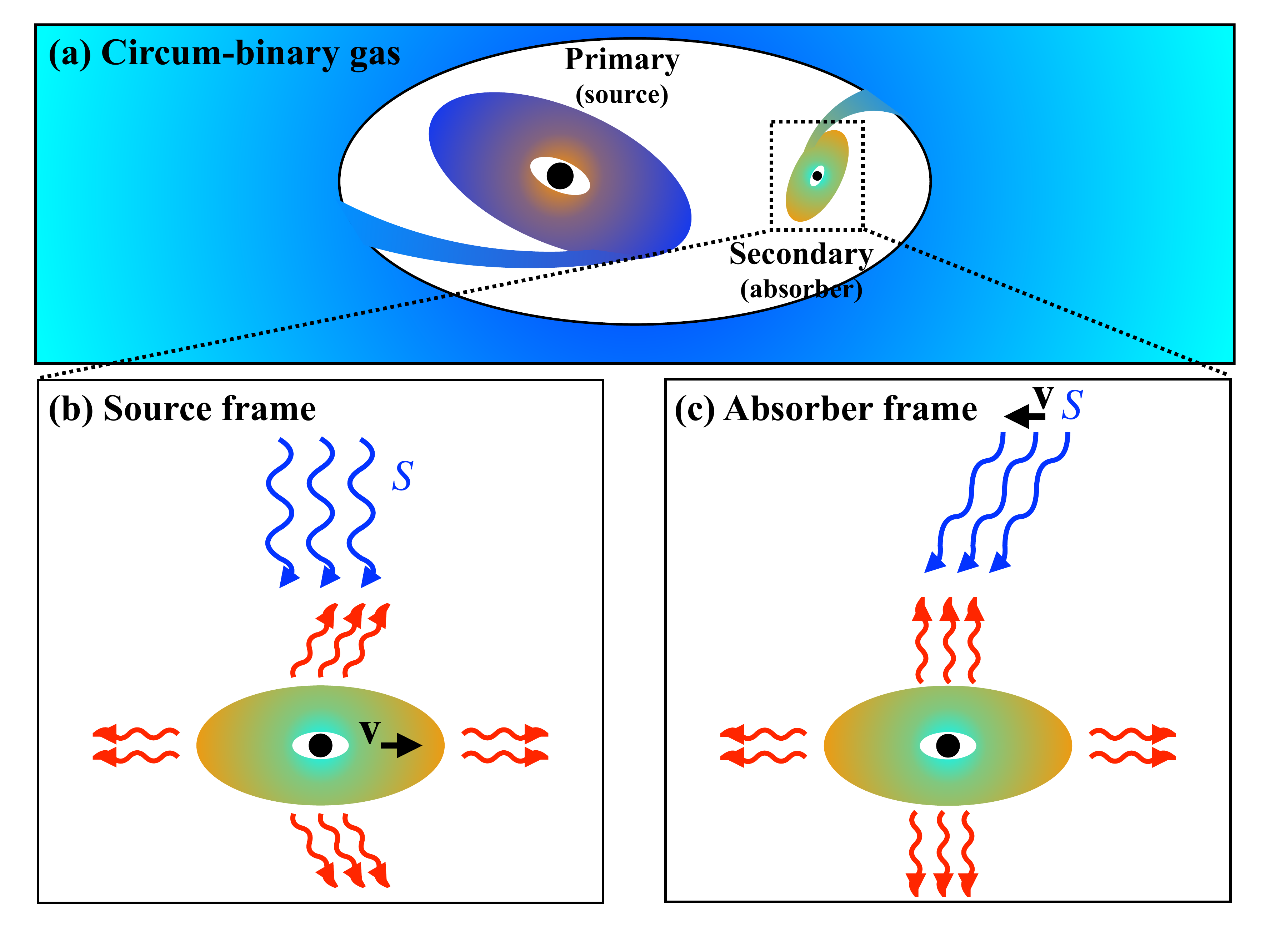}
\caption{Physical picture of the PR drag. (a) A SMBHB embedded in a gaseous environment
and both BHs are accreting. (b) In the rest frame of the light
	source (the circum-primary disk), the secondary BH is moving at a velocity of $v$. The light from the circum-secondary disk is beamed in the
	direction of motion, causing a linear-momentum loss. To conserve linear
	momentum, the circum-secondary disk and the embedded SMBH decelerates.
	(c) In the frame comoving
	with the absorber (the circum-secondary disk), the source is moving. The emission of the circum-secondary disk is now
	isotropic, but the light rays from the source is inclined
	due to the beaming effect. This inclined irradiation causes
	the circum-secondary disk and the embedded SMBH to drift relative to
	the comoving frame.
} \label{fig:pic} \end{center}
\end{figure}

We consider a SMBHB embedded in an gaseous environment, and both the primary
and secondary BHs are accreting gas.  The configuration is illustrated in
panel (a) of Figure~\ref{fig:pic}.  We note that the two accretion disks,
namely the circum-primary and the circum-secondary disks, are not necessarily
coplanar or aligned with the orbital plane of the SMBHB, as is shown in the
numerical simulations of accreting SMBHB systems
\citep[][]{dotti10,nixon13,gerosa15,goicovic16,takakuwa17} as well as the
observations of the circum-stellar disks in binary protostars
\citep{takakuwa17}.  Therefore, both accretion disks could be irradiated by the
companion.  For simplicity, we consider the circum-primary disk as the
light source and the circum-secondary one as the absorber. In the following, we
study the PR drag exerted on the secondary BH.

The physical picture of the PR drag is shown in the lower two panels of
Figure~\ref{fig:pic}. (b) In the rest frame of the irradiating source (the
circum-primary disk), the secondary SMBH is moving in a direction perpendicular
to the light rays from the source. An absorption of the light by the
circum-secondary accretion disk does not change this perpendicular velocity.
Meanwhile, the circum-secondary disk is re-emitting more light in the direction
of the motion.  This beamed emission carries momentum and, by the law of
linear-momentum conservation, the circum-secondary disk and the embedded SMBH
should recoil in the opposite direction. This recoiling effectively causes the
PR drag.  We note that the irradiation by an external source is crucial to the
PR drag. Without it, although the emitted light from the small body is still
beamed, it does not slow down the moving body because the light also takes away
the rest mass, compensating the loss of the linear momentum \citep[pointed out
by][]{robertson1937}. (c) The same conclusion can be drawn in the rest frame of
the absorber, i.e., the circum-secondary disk. In this frame, the emission from
the circum-secondary disk is isotropic but the light rays from the source
becomes inclined because the source is now moving.  The irradiation imposes a
pressure on the circum-secondary disk and because of the inclination, the
pressure has a component pointing in the direction of the motion of the source.
This component forces the secondary disk to drift with respect to the comoving
frame. This drift is equivalent to the recoiling effect seen in the source
frame.

We now go back to the source frame and the drag force due to the PR effect can be calculated with
$Sv/c^2$ \citep{robertson1937}, where, in our scenario, $v$ is the orbital
velocity of the secondary SMBH, $c$ is the speed of light, and $S$ denotes the
energy flux that is incident on the circum-secondary disk.
The above equation assumes that all the energy flux is absorbed by the disk,
which is generally true for optically thick accretion disks.  As a result of
the PR drag, the secondary BH decelerates and migrates towards the primary 
on a timescale of
\begin{align} 
	T_{\rm PR}=\frac{mv}{S\,(v/c^2)}=\frac{mc^2}{S}. \label{eq:Tpr1}
\end{align}

To derive the value of $T_{\rm PR}$, we first express the energy flux using
$S=\xi Lr^2/(4R^2)$, where $L$ is the bolometric luminosity of the source, $r$
denotes the radius of the circum-secondary disk, and $R$ is the distance
between the primary and secondary BHs.  The coefficient $\xi$ characterizes the
cross section of the circum-secondary disk in the radiation field, and it is a
function of the inclination of the disk relative to the light rays from the
source.  It is of order unity if the disks are misaligned. Even when the two
disks are coplanar, the value of $\xi$ does not vanish because accretion disks
are not infinitely thin. The luminosity can be further written as $L=\eta
L_{\rm Edd}$, where $L_{\rm Edd}$ is the Eddington luminosity and $\eta$ is the
``Eddington ratio''.  If the primary BH has a mass of $M$, the Eddington
luminosity is $L_{\rm Edd}\simeq1.26\times10^{38}(M/M_\odot)\,{\rm
erg\,s^{-1}}$.  For the size of the circum-secondary disk ($r$), we notice that
earlier numerical simulations found that it is comparable to the Roche radius
$R_L:=Rq^{1/3}(1+q)^{-1/3}$ \citep{lin76,artymowicz94,mosta19}, where $q:=m/M$
is the mass ratio between the secondary BH and the primary one ($q\le 1$ by
definition).

In the derivation we assumed that the radiation field in the source frame is
isotropic. It is known that the radiation from AGN accretion disks could be
collimated when $\eta$ is close to or exceeds the Eddington limit. In this
case, if the radiation directly impacts the circum-secondary disk, the momentum
flux would be greater than what we have estimated above. Moreover, accretion
disks with high Eddington ratios also produce outflows or jets. These
structures also carry momentum.  If they strike the circum-secondary disk, the
interaction could induce an additional drag force which is similar to the PR
drag, only different in the sense that it is caused by the mass-momentum flux.
We do not consider these additional effects in this work. Therefore, our PR
timescale should be regarded as a upper limit.

With these considerations, we can rewrite Equation~(\ref{eq:Tpr1}) as
\begin{align}
	T_{\rm PR}=\frac{4mc^2}{L\,(r^2/R^2)}
	\simeq\frac{4q^{1/3}(1+q)^{2/3}}{\xi\,\eta}\frac{Mc^2}{L_{\rm Edd}}.\label{eq:inter}
\end{align}
Note that $M\,c^2/L_{\rm Edd}$ is a constant independent of the BH mass or the
distance between the two BHs. As a result,
\begin{equation}
	T_{\rm PR}\simeq1.8\times10^{9}\,\xi^{-1}\eta^{-1}q^{1/3}(1+q)^{2/3}\,{\rm years}. \label{eq:Tpr2}
\end{equation}
Therefore, despite the uncertainties in the hydrodynamics, the PR timescale is
well determined by three parameters, namely, the mass ratio of the two BHs
($q$), the Eddington ration for the primary BH ($\eta$), and the relative
inclination of the two accretion disks ($\xi$).

\begin{figure}
\begin{center}
\includegraphics[clip,width=0.5\textwidth]{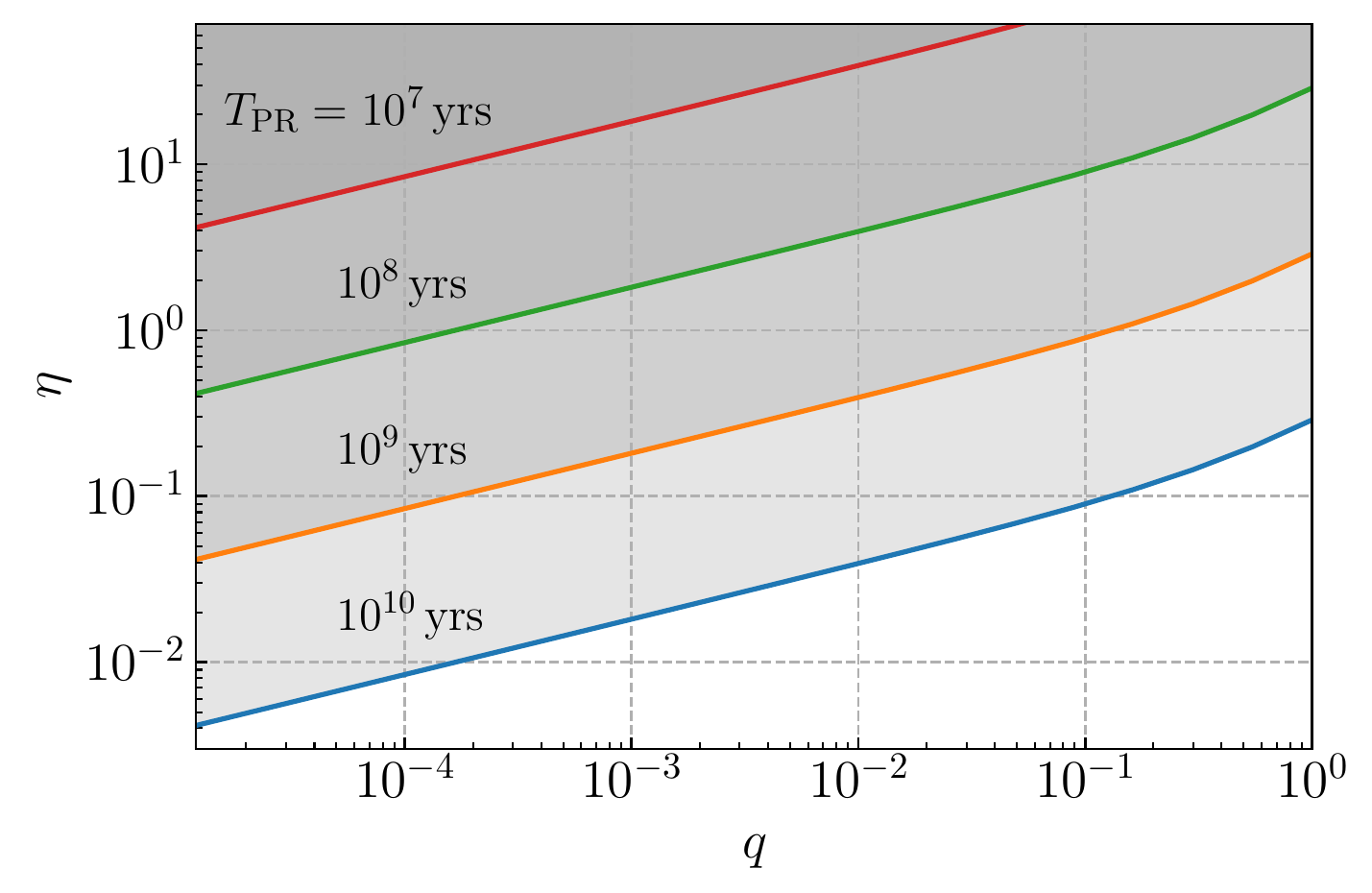}
	\caption{Dependence of the PR timescale ($T_{\rm PR}$) on the mass ratio ($q$) 
	of the SMBHB and the Eddington ratio ($\eta$) of the accreting primary
	BH. The curves represent the contours of constant $T_{\rm PR}$.} \label{fig:eta}
\end{center}
\end{figure}

Assuming that the inclination angle between the two accretion disks is large
($\xi\sim1$), we show in Figure~\ref{fig:eta} the dependence of $T_{\rm PR}$ on
the other two parameters.  Interestingly, $T_{\rm PR}$ is shorter than the
Hubble time ($10^{10}$ years) in a large fraction of the parameter space.  In
particular, for equal-mass binaries ($q\simeq1$), coalescing within a Hubble
time requires that the Eddington ratio is greater than about $0.3$. For unequal
binaries, e.g., $q\la0.1$, the requirement becomes $\eta\ga0.1(q/0.1)^{1/3}$.

\section{Compare with other timescales}

To understand the relative importance of the PR drag, we compare $T_{\rm PR}$
with the other timescales related to the formation and evolution of SMBHBs.

(1) We first consider the gravitational-wave radiation
timescale. We calculate it with
\begin{align}
	T_{\rm gw}&=\frac{5}{64}\frac{R^4c^5}{G^3(M+m)Mm}\\
	&\simeq2.3\times10^{7}q^{-1}(1+q)^{-1}M_8^{-3} \left(\frac{R}{0.01\,{\rm pc}}\right)^{4}\,{\rm yrs}  
\end{align}
assuming circular orbits \citep{peters63} and denoting $M/(10^8\,M_\odot)$
with $M_8$. The condition $T_{\rm PR}<T_{\rm gw}$ could be satisfied when
\begin{equation}
	R\ga 0.03\xi^{-1/4}\eta^{-1/4}q^{1/3}(1+q)^{5/12}M_8^{3/4}\,{\rm pc}.
\end{equation}
Therefore, the PR drag predominates at relatively large binary separation.

(2) Dynamical friction against the stellar background could also shrink the
orbit of a SMBHB. Following \cite{BT08}, we calculate the dynamical-friction
timescale with $T_{\rm df}=\sigma_*^3/(4\pi G^2\rho_*m\ln\Lambda)$, where
$\rho_*$ is the mass density of the stellar background, $\ln\Lambda\simeq6$ is
the Coulomb logarithm, and in deriving the above equation we have assumed that
the gravitational potential of the primary BH predominates so that the orbital
velocity of the secondary, $v$, is comparable to the velocity dispersion of the
background stars, $\sigma_*$.  Now we evaluate this $T_{\rm df}$ within the
gravitational influence radius of the primary BH, $R_{\rm inf}\simeq
GM/\sigma_*^2$, within which the gravity of the primary BH predominates.
We have to consider this restriction because
the PR timescales derived above become invalid outside the influence radius.  
Noticing that (i) empirically
$\sigma_*(R_{\rm inf})=200M_8^{1/4}\,{\rm km\,s^{-1}}$ \citep{tremaine02},
(ii) $\sigma_*^2(R)\simeq GM/R$ within the influence radius,
(iii) the stellar mass enclosed in
the binary orbit is about $M_*\sim4\pi\rho_* R^3/3\propto R^{3-\gamma}$ where
$\gamma$ denotes the power-law index of the density profile ($\rho_*\propto R^{-\gamma}$), 
and (iv) $M_*(R_{\rm
inf})\sim M$ according to the definition of the influence radius, we deduce
that
\begin{align}
	T_{\rm df}&\sim\frac{1}{3\ln\Lambda}\left(\frac{M^2}{mM_*}\right)
	\left(\frac{R}{\sigma_*}\right)
	\simeq\frac{3000M_8^{1/4}}{q}\left(\frac{R}{R_{\rm inf}}\right)^{\gamma-3/2}\,{\rm yrs}.
\end{align}
Comparing it with Equation~(\ref{eq:Tpr2}), we find that $T_{\rm PR}<T_{\rm
df}$ when $q\lesssim5\times10^{-5}(\xi\eta)^{3/4}M_8^{3/16}(R/R_{\rm inf})^{3\gamma/4-9/8}$. 
This result suggests that the PR drag is more important than dynamical 
friction for unequal binaries.

(3) Salpeter timescale ($T_{\rm Sal}$) characterizes how fast an accreting
	body increases its mass by one e-folding. It can be shown that $T_{\rm
		Sal}$ is closely related to $T_{\rm PR}$. 
Given a radiative efficiency of $\epsilon$ for an accretion disk ($\epsilon\sim0.1$),
the Salpeter timescale can be calculated with 
\begin{align}
T_{\rm Sal}&:=\frac{\epsilon Mc^2}{(1-\epsilon)L}
=\frac{\epsilon}{(1-\epsilon)\,\eta}\frac{Mc^2}{L_{\rm Edd}}\\
&\simeq 4.53\times10^8\epsilon\,\eta^{-1}(1-\epsilon)^{-1}\,{\rm years}. \label{eq:Tsal}
\end{align}
Using the relationship between $T_{\rm Sal}$ and $Mc^2/L_{\rm Edd}$, we can
rewrite Equation~(\ref{eq:inter}) as
\begin{equation}\label{eq:TPRTSAL}
T_{\rm PR}\simeq\frac{4q^{1/3}(1+q)^{2/3}(1-\epsilon)}{\xi\,\epsilon}T_{\rm Sal}.
\end{equation}
The two timescales become comparable when $q\simeq q_{\rm
cri}:=(\xi\epsilon/4)^3$. For $\xi=1$ and $\epsilon=0.1$, we find
that $q_{\rm cri} \simeq1.6\times10^{-5}$. 

When $q<q_{\rm cri}$, the PR timescale becomes shorter than the Salpeter
timescale. In this case, 
a SMBH could clear away the surrounding small BHs before it grows by
one e-folding.

For a SMBHB with  $q>q_{\rm cri}$, the PR timescale is longer than the Salpeter
timescale, i.e., $T_{\rm PR}>T_{\rm Sal}$. In this case, to drive the SMBHB to
coalescence, the primary BH must grow by more than one e-folding.  This
scenario applies to those BHs in the early universe, where they have to accrete
enough gas to grow from a mass of $10^2-10^5\,M_\odot$ to the current
$10^6-10^9\,M_\odot$ \citep{volonteri10}.  The growth, in fact, amounts to nine
e-foldings.  Therefore, if we take $T_{\rm PR}<9T_{\rm Sal}$ as the criterion
for binary coalescence, adopting the assumption that $\epsilon=0.1$ and
$\xi=1$, we find that $q\la0.016$.

(4) Duty cycle ($T_D$) is another important timescale.  Conventionally, it is
designed to characterizes the lifetimes of AGNs. In our problem, it provides an
estimation of the total duration the accretion episodes of a SMBHB.  Only when
$T_{\rm PR}\la T_{D}$ is the PR drag efficient enough to affect the dynamical
evolution of the binary.  Observations of luminous AGNs suggest that $T_{D}$ is
a decreasing function of $\eta$ \citep{hopkins09,shankar09}.  For
$0.1\la\eta\le1$, $T_{D}$ is typically $10^8$ years, and for
$0.01\la\eta\la0.1$, $T_{D}$ increases to $10^9$ years.  These results in
general agree with the Salpeter timescale as is derived in
Equation~(\ref{eq:Tsal}).  Using these values for the duty cycle, we find that
the condition $T_{\rm PR}< T_{D}$ is satisfied when $q\la1.7\times10^{-4}$
according to Equation~(\ref{eq:Tpr2}). We note that this requirement applies
mainly to those SMBHBs in the local universe, because the duty cycles used in
our analysis are derived based on relatively low-redshift AGNs. 

(5) Small objects on inclined orbits with respect to an accretion disk could be
ground down into the disk due to the mutual collisions \citep{syer91}. We adopt
the formula in \cite{ivanov99} and calculate the ``ground-down'' timescale with
$T_{\rm gd}=m /(\Sigma A \Omega)$, where $\Sigma$ is the surface density of the
accretion disk at the point of collision, $\Omega=v/R$ is the angular velocity
of the secondary BH, and $A$ is the effective cross section of collision.  The
above equation is derived in the approximation that the relative velocity of
the BH-disk collision is of the order of $v$.

To compare $T_{\rm gd}$ with $T_{\rm PR}$, we
first calculate the collisional cross section with $A=\pi(Gm/v^2)^2$. We 
also note that in the standard accretion-disk model, 
$\Sigma$ is related to the accretion rate $\dot{M}$ as $\dot{M}=3\pi \nu\Sigma$,
where $\nu$ is the viscosity and it is related to the viscosity parameter $\alpha$
and the disk scale hight $H$ as $\nu=\alpha \Omega H^2$ \citep{frank02}.
From these relations, we find that
\begin{equation}
	T_{\rm gd}\simeq3\alpha h^2 q^{-1}(1-\epsilon) T_{\rm sal},
\end{equation}
where $h:=H/R$ is the aspect ratio of the disk. It is now clear that when
$q\lesssim 8\times10^{-4}$, the PR timescale would be shorter than the
ground-down timescale, if we adopt the typical parameters
$\alpha=h=\epsilon=0.1$ and $\xi=1$. In this case, the binary would have
shrunk significantly due to the PR drag before its orbit becomes coplanar the
accretion disk.

(6) Even after the secondary BH has been ground down into the accretion disk, the
PR drag may not vanish completely because the disk could be warped, e.g.,
 due to the Bardeen-Petterson effect \citep{bardeen75}.
It is well known that an embedded secondary would excite density waves in the
accretion disk \citep{goldreich80}. If the mass of the secondary is small, the
disk surface density is not significantly perturbed and the interaction between
the secondary and the disk can be calculated in a linear approximation. The
interaction would result in a radial migration of the secondary, which is known
as the type-I migration. The migration timescale can be calculated with
\begin{equation}
	T_I=\frac{fh^2M}{q\Sigma R^2\Omega}\simeq3\pi\alpha h^4fq^{-1}(1-\epsilon)T_{\rm Sal},
\end{equation}
where $f$ is a parameter depending on the temperature and density profiles of
the disk near the secondary's orbit \citep[e.g.][]{Paardekooper11}, and we have applied the relation
$\dot{M}=3\pi\nu\Sigma$ in the second equation. Using our fiducial parameters
and Equation~(\ref{eq:TPRTSAL}), we find that $T_{\rm PR}<T_I$ when
$q\lesssim6\times10^{-5}$.  It is worth noting that the direction of the type-I
migration could be inward or outward depending on the sign of $f$, which in
turn depends on the exact temperature and density profiles of the accretion
disk. Meanwhile, the PR drag always leads to an inward drift.

A small, embedded secondary BH could also accrete from the accretion disk.
The increase in mass leads to an inward migration
of the secondary because of the conservation of angular momentum.  
To estimate the corresponding timescale, we calculate the deceleration 
due to accretion with $\dot{v}=\pi \rho R_B^2 \Delta V^2/m$, where $\rho\sim \Sigma/(2H)$
is the surrounding gas density, $\Delta V$ is the relative velocity between the secondary and the surrounding gas, which is of the order of the sound speed $c_s$, 
and $R_B=Gm/\Delta V^2$ is the Bondi radius. Using the condition 
$c_s=\Omega H$
for hydrostatic equilibrium, we find that the migration timescale is
\begin{equation}
	T_M:=v/|\dot{v}|\simeq6\alpha h^5 q^{-1} (1-\epsilon) T_{\rm Sal}.
\end{equation}
Compared to $T_{I}$, $T_M$ is more sensitive to $h$, so that the condition
$T_{\rm PR}<T_M$ requires an even smaller $q$, i.e., $q\lesssim8\times10^{-6}$
in our fiducial model.

If the secondary is massive enough, an annular gap could be opened in the disk
around the orbit of the secondary. In this case, the secondary will be 
locked in the gap and migrate on a timescale correlated with the viscous timescale
of the disk, $t_{\rm vis}=2R^2/(3\nu)$. Following \cite{lin86}, we calculate
the timescale of this  type-II migration with $T_{II}=t_{\rm vis}(m/M_d)$,
where $M_d\simeq \pi R^2 \Sigma$ is the disk mass enclosed in the orbit of the 
secondary and we have assumed that $M_d\ll m$. For a standard thin disk, we find 
that
\begin{equation}
	T_{II}\simeq2q(1-\epsilon)T_{\rm Sal},
\end{equation}
and, in fact, it is shorter than $T_{\rm PR}$ for any $q$ if we adopt
the fiducial parameters of $\eta=\epsilon=0.1$. This result
indicates that if the secondary becomes massive enough to open a gap in the
accretion disk, the later evolution would be dominated by type-II migration and
the PR drag is relatively unimportant.

\section{Discussion}

In our problem, the object receiving the PR drag is quite different from
a solid dust particle. Nevertheless we calculated the drag using the formula
derived for dust particles. Such a simplification deserves justification. 

(i) In our problem the drag force is imposed directly on the circume-secondary
accretion disk. However, the gravitational coupling between the disk and the
secondary BH is so strong that the drag force can be imparted to the BH almost
immediately. One can see this by calculating the dynamical timescale of the
circum-secondary disk, which characterizes how quickly the accretion disk
responds to a perturbation. We can calculate it with $2\pi(Gm/R_L^3)^{-1/2}$,
which equals $2\pi(GM/R^3)^{-1/2}$. The dynamical timescale is the longest when
the binary is at a distance of $R_{\rm inf}$, but even in this case it is
approximately $R_{\rm inf}/\sigma_*\simeq5\times10^3M_8^{1/4}$ years.  We see
that the timescale is indeed much shorter than the PR timescale. Therefore, any
offset between the BH and the surrounding accretion disk will be damped
relatively quickly.

(ii) Our emitter, i.e., the accretion disk, is rotating but it should not
significantly affect the calculation of the drag force. This is so because the
rotation velocity is typically $v_c\sim(Gm/R_L)^{1/2}$, and we find that
$v_c/v\sim q^{1/3}$.  Therefore, when $q\ll1$, most of the gas in the
circum-secondary disk is rotating at a velocity smaller than the orbital
velocity, so the rotation can be neglected.  Even when $q\sim1$ so that
$v_c\sim v$, the previous calculation of the PR drag force is approximately
correct, because an axisymmetric rotation does not break the symmetry of the re-emission and
hence does not contribute to the PR drag.  The orbital motion induces asymmetry
due to aberration, and hence is the main source of the PR drag.

(iii) Our Figure~\ref{fig:eta} includes a region where the luminosity is
super-Eddington ($\eta>1$).  Such large luminosity should not destroy the
circum-secondary disk by blowing it off.  This is because the
conventional Eddington luminosity, which is used in this work, is derived
assuming an opacity of $\kappa_e\simeq0.4\, {\rm cm^2\,g^{-1}}$, dominated by electron
scattering. However,  accretion disks are normally optically thick, so that
$\kappa_e\Sigma\gg1$. As a result, the effective opacity is $1/\Sigma$ and it
is much smaller than $\kappa_e$. Since Eddington luminosity is inversely proportional
to the opacity, the corresponding effective Eddington
luminosity, to blow away the circum-secondary accretion disk, is much higher
than the conventional one.

(iv) Since the dynamical friction timescale $T_{\rm df}$ is inversely
proportional to $q$, it seems that very small BH cannot come from outside the
influence radius and be delivered to the vicinity of a SMBH. However, small BH
can form in situ, as the remnants of massive stars, or be brought in by massive
star clusters. How fast these channels populate the galactic nuclei with small
BH is out of the scope of this work and deserves further investigation.

\section{Conclusion}

In this Letter, we investigate the impact of a new drag force, induced by the
Poynting-Robertson effect, on the dynamical evolution of SMBHBs.  We find that
for a mass ratio of $q\lesssim$ a few$\times10^{-5}$, the PR drag could predominate the
dynamical evolution and lead to a fast coalescence of the BHs.  The relevant
systems include stellar-mass BHs of ${\cal O}(10)\,M_\odot$ around
$10^6-10^7M_\odot$ SMBHs (the mergers are known as the ``extreme-mass-ratio
inspirals''), as well as intermediate-massive BHs ($10^3-10^{5}M_\odot$)
around  $10^8-10^{10}M_\odot$ SMBHs.  Unlike the dynamical-friction or
type-I/II migration timescales, which are sensitive to the properties of the
stellar and gas distribution around SMBHs and hence are uncertain, the PR
timescale is determined by fewer parameters, essentially only the mass ratio
$q$ and the Eddington ratio $\eta$. Our work made it possible to implement the
PR drag in the future hydrodynamic and cosmological simulations so that we can
better understand the evolution of unequal SMBHs in galactic nuclei.

\acknowledgments

This project is supported by the National Science Foundation of China (grants
No. 11721303, 11873022, and 11991053).

\bibliographystyle{astroads.bst}

\end{document}